# In-memory phononic learning toward cognitive mechanical intelligence

Yuning Zhang, K. W. Wang*

Department of Mechanical Engineering, University of Michigan, Ann Arbor, MI 48109, USA.

* **Corresponding author:** K. W. Wang (aka: Kon-Well Wang)

Email:  kwwang@umich.edu,  Phone number: (734)764-8464

Address: 2468 G.G. Brown Building, Ann Arbor, MI 48109, USA

**Abstract**

Modern autonomous systems are driving the critical need for the next-generation of adaptive materials and structures with embodied intelligence, i.e., the embodiment of memory, perception, learning, and decision-making within the mechanical domain. A fundamental challenge is the seamless and efficient integration of memory with information processing in a physically interpretable way that enables cognitive learning and decision-making under uncertainty. Prevailing paradigms, from intricate logic cascades to black-box morphological computing or physical neural networks, are seriously limited by trade-offs among efficiency, scalability, interpretability, transparency, and the reliance on additional electronics. Here, we introduce in-memory phononic learning, a paradigm-shifting framework that unifies non-volatile mechanical memory with wave-dynamics-based perception within a phononic metastructure. Our system encodes spatial information into stable structural states as mechanical memory that directly programs its elastic wave-propagation landscape. This memory/wave-dynamics coupling enables effective sensory perception, decomposing complex patterns into informative geometric features through frequency-selective wave localization. Learning is created by optimizing input waveforms to selectively probe these features for memory-pattern classification, with decisions inferred directly from the output wave energy, thereby completing the entire information loop mechanically through an efficient and physically transparent mechanism without hidden architectures or electronics. We experimentally and numerically investigated exemplar high-accuracy classification across diverse tasks, from geometric shapes to handwritten digits. This work transcends the paradigm of "materials that compute" to "cognitive matter" capable of interpreting the dynamic environments, paving the way for future intelligent structural-material systems with low power consumption, more direct interaction with surroundings, and enhanced cybersecurity and resilience in harsh conditions.

**Introduction**

Mechanical intelligence (MI)[1–5]—the concept of designing intelligence within the mechanical domain of materials and structures—has recently emerged as a direction to enable autonomous engineering systems with high performance, efficiency, and resilience. Instead of outsourcing entirely to electronics and computers, MI envisions materials that store, process, and perceive information, and adapt to environmental inputs through their intrinsic mechanical responses, thereby reducing latency and energy overheads while improving robustness in extreme conditions and resistance to cyber threats. These unique features and advantages make MI particularly critical for the realization of emerging complex engineering systems demanding advanced autonomy, such as smart robotics, adaptive infrastructure, automated vehicles, and personal wearables.

The foundation of MI lies in mechanical information processing and learning[6–9], which requires a memory module to retain knowledge from environments, a computing module to process, interpret, and perceive information, and a physical communication interface that bridges the stored memory with computation. A widely adopted strategy emulates the digital computing paradigm by abstracting mechanical bits from multistable structures for non-volatile memory storage and implementing logic operations via carefully designed quasi-static state transitions and snap-through interactions[10–19]. However, scaling such logic-based architectures to high-level computing tasks—especially cognitive learning and adaptation—requires sequential cascading of numerous basic logic components via complex mechanical coupling, where information is transferred between modules through slow quasi-static signal propagation, limiting both operational efficiency and architectural scalability.

Learning naturally demands substantially greater computing power, requiring systems to store and reason information, perceive features, and make decisions under uncertain environments. Inspired by artificial or biological neural architectures, morphological computing or physical neural networks—including mechanical[20–24] or wave-based neural networks[25–29] and physical reservoir computing (PRC)[30–35]—has emerged to pursue learning using architected metamaterials. This approach supports higher-level cognitive tasks, such as machine-learning regression and classification tasks, through distributed physical transformations while avoiding explicit logic cascades. Yet, these systems often depend either on sophisticated structural optimization for task-specific training, or on complex nonlinear and dynamic reservoirs requiring external digital readouts. Critically, learned input-output mappings emerge from opaque black-box correlations within disordered and intractable structural responses, undermining reliability in scenarios demanding transparent reasoning and preventing the establishment of predictive and transferable design rules—factors crucial for robustly deploying MI in real-world applications. What is missing, therefore, is a unified, efficient, and physically transparent framework that seamlessly couples memory with other elements of intelligence within the mechanical domain, in which stored information natively drives computation and perception, enabling learning and decision-making through interpretable and predictable mechanisms.

The goal of our research is to advance the state of the art by bridging the abovementioned missing gap via creating a fundamentally different and transformative framework of *in-memory phononic learning*. This approach harness elastic wave propagation in architected phononic metastructures as an efficient and mechanistically explicit interface that directly connects memory and the information processing for learning. Phononic metastructures[36,37] are an ideal medium for this role; through rich wave dispersion, interference, and resonance, they can perform complex analog computations at the speed of sound while supporting parallelism across frequencies and modes[38–40], where these features will be leveraged to interact with and perceive information stored in non-volatile mechanical memory.

To realize the new concept and evaluate our hypotheses, we develop an innovative multistable phononic metastructure testbed that encodes complex spatial information from environmental

inputs—such as payload or deformation patterns—into reconfigurable, stable metastructure states as non-volatile mechanical memory (Figure 1A). Wave-based mechanical perception and cognitive learning capabilities are then investigated and demonstrated through exemplar pattern-recognition tasks, where structured elastic wave propagation is harnessed to perform in-situ classification of these encoded memory patterns (Figure 1B). Specifically, the innovation of this testbed, detailed in the Results section, is that it functionally separates memory state switching from wave propagation while coupling them through controllable, state-dependent interactions. It enables the testbed to concurrently achieve most effective wave transmission and state switching via different materials in different layers. As a consequence, the metastructure memory directly and deterministically programs the system's elastic wave-propagation landscape, transforming the stored information into rich vibrational modes via the formulation of elastic waveguides. In-situ perception occurs in the frequency domain by harnessing vibrational localization modes: under different harmonic excitations, the memory-defined geometry localizes wave energy into distinct, frequency-selective local waveguides, for example, along specific edges or corners, which acts as spatial phononic "filters" to extract key features of the stored information. We then establish a phononic learning framework to classify encoded memory patterns from the perceived knowledge by applying input excitation waveforms in the forms of linearly weighted harmonic signals (Figure 1B). Instead of optimizing the structural geometry, our training process optimizes these input waveforms to selectively activate or suppress feature-modes associated with specific memory pattern classes, thereby redistributing wave energy across the output channels. The classification decision is then inferred directly from the output energy distribution across channels, representing a physical analog of prediction probabilities. In principle, this architecture is inspired by but very different to that of PRC, by keeping learning and decision in the phononic domain without relying on an external electronic readout layer. It creates a wave-physics-informed analog neural network, where individual frequency responses serve as fixed feature-extraction layers operating in parallel across frequencies (i.e., constituting a physical reservoir), while the superposition principle allows the input waveforms to act as a trainable linear output layer to combine perceived features and perform classification (Figure 1C). This framework completes the entire information-processing chain—from memory encoding to perceptual feature extraction to learned decision—within a single, fixed mechanical substrate. The process is efficient, inherently parallel, and physically interpretable, as each step is governed by predictable and traceable memory-wave interactions in a structured medium, avoiding the black-box nature of disordered morphological systems. Furthermore, the framework naturally exhibits versatile computing power for multi-tasking, where different tasks can be executed with the same structure by simply updating the input waveforms. We experimentally and numerically validate this in-memory phononic learning framework by achieving high-accuracy recognition across a series of 2D spatial pattern classification tasks. By unifying memory, perception, and learning through a physically interpretable and transparent mechanism, we elevate the ambition from "materials that compute" to the next level – achieving "cognitive materials" that learn and understand through their inherent wave physics, providing a foundation for deployable and robust mechanically intelligent matter that can interpret, reason, and adapt to dynamic, unpredictable, and data-rich environments.

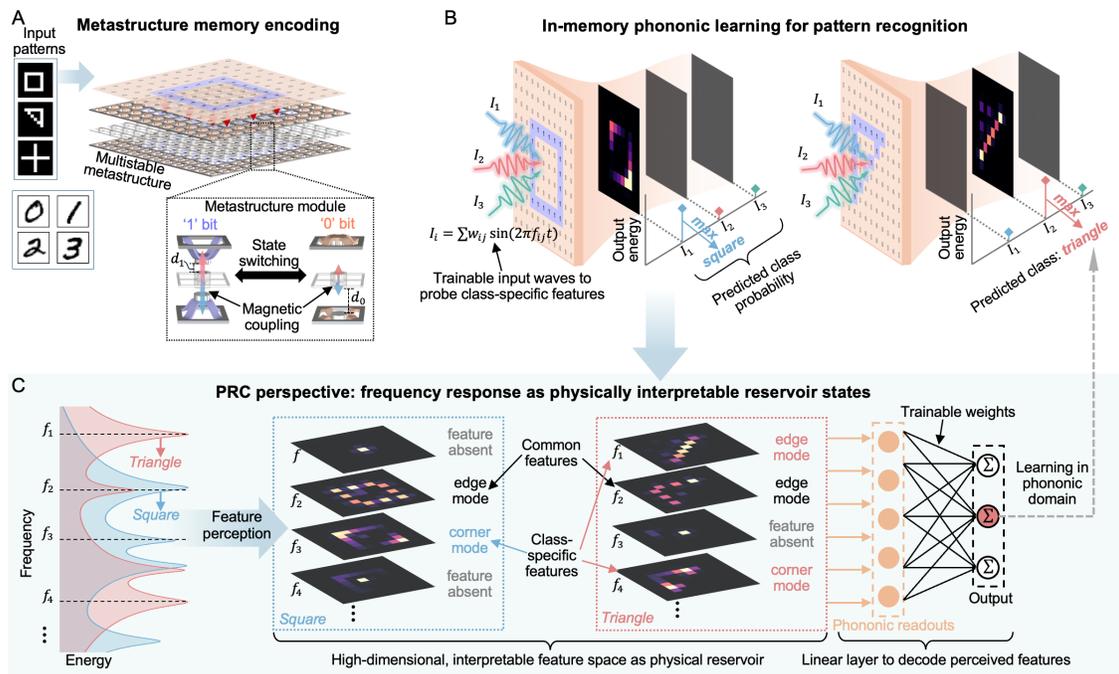

**Figure 1. In-memory phononic learning framework.** (A) Memory encoding in a multistable phononic metastructure. The two symmetric stable states of each unit cell module are abstracted as mechanical bits '0' and '1' and can be independently programmed to write spatial information as non-volatile mechanical memory. (B) In-memory phononic learning for pattern recognition. Trainable input waveforms (linearly weighted superpositions of harmonic tones) probe class-specific, frequency-selective vibrational localization modes. The classification decision is inferred directly from the output energy per input channel with the energy distribution across channels as a physical analog of class probabilities. (C) Conceptual equivalence to physical reservoir computing: treating the wave physics as a physical analog neural network. The metastructure's frequency responses form a fixed, high-dimensional, and interpretable physical reservoir in which diverse localization modes perceive class-specific geometric features, and the trainable input waves act as a linear readout layer, combining extracted features for classification outputs.

**Results**

**Metastructure design for programmable non-volatile mechanical memory and wave physics**. As illustrated in Figure 2A, a three-layer multistable phononic metastructure is built that functionally separates memory encoding from wave propagation while coupling them through controllable, state-dependent interactions. It comprises an inner phononic plate, sandwiched by two outer multistable layers that provide non-volatile mechanical memory. The inner plate is a locally resonant phononic metamaterial consisting of periodic square units, each containing a central resonator mass connected to a square frame through four thin beams (Figure 2B). This layer is purposefully fabricated from a low-damping material to preserve high-fidelity wave responses as needed for wave-based information processing. The top and bottom layers are arrays of curved-beam bistable elements mounted in a rigid frame, made from soft materials to facilitate effective state-switching for mechanical memory encoding and storage. Each opposing bistable pair is aligned with one inner-layer resonator so that the three pieces form a single cell multistable module (Figure 2C), which functions as a mechanical memory bit that can be independently programmed to write information. To physically link the stored memory to wave physics, permanent magnet disks are affixed to the center of each outer bistable element and to each inner-layer resonator mass. The magnets are axially magnetized and positioned to generate repulsive forces between the inner resonator and its two outer counterparts. State-switching of the bistable pair changes the inter-magnet distance and thus modulates the local coupling. Here, we consider the two symmetric stable states for each module as mechanical bits '0' and '1', where bit '0' means both outer bistable elements are switched outward with an inter-magnet distance of $d_0 = 1.7$ cm, resulting in low magnetic coupling, and bit '1' means both are switched inward with reduced inter-magnet distance of $d_1 = 0.6$ cm, strengthening the magnetic coupling (see dashed box in Figure 1A). Figure 2E shows our experimental prototype of an 11-by-11 phononic metastructure (see Supporting Information, Section S1 for details of the experimental fabrication and design parameters). We experimentally validate the multistability of the testbed via quasi-static tensile testing on individual modules and demonstrate the programmable, non-volatile mechanical memory for information encoding at the array level by manually switching selected modules to write spatial patterns into the testbed (see Section S2 for details of the state-switching behavior of the metastructure module, and examples of mechanical information encoding into the testbed).

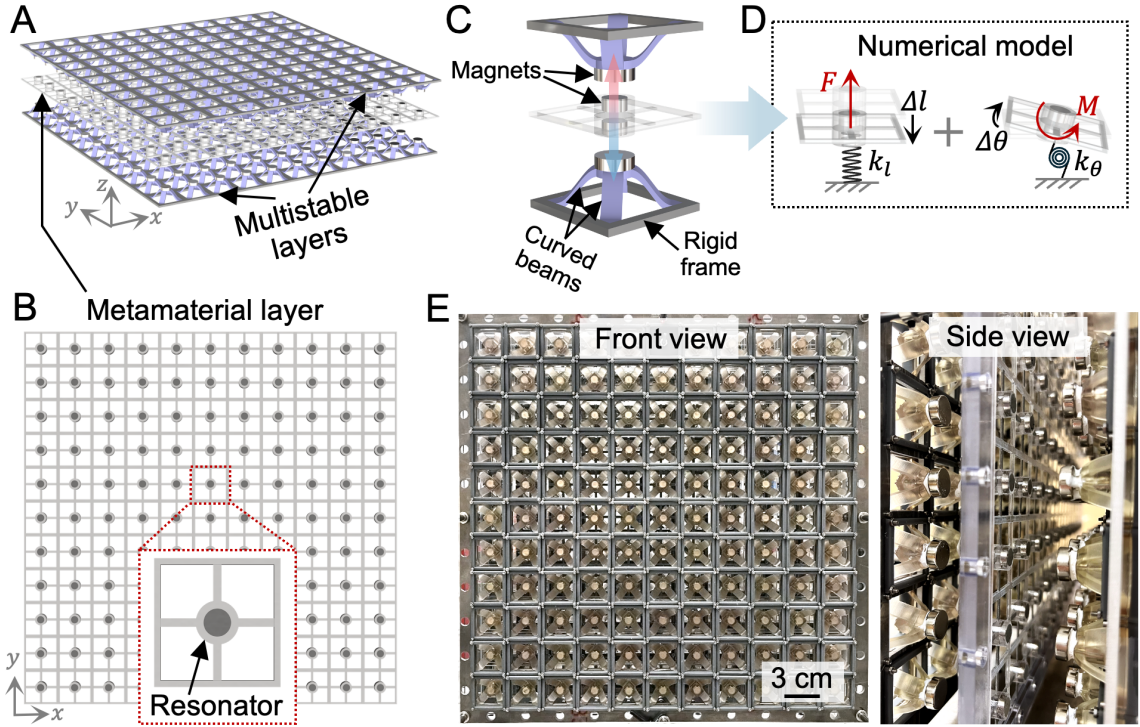

**Figure 2. Multistable phononic metastructure design and modeling.** (A) Schematics of the three-layer phononic metastructure, containing a low-damping inner phononic metamaterial plate sandwiched by two multistable outer layers that encode non-volatile mechanical memory. (B) Inner phononic metamaterial layer consists of periodic square units with side length $a = 3$ cm, each with a central resonator mass. (C) A multistable unit cell module with two outer curved-beam bistable elements mounted in a rigid frame and an inner local resonant unit. Magnetized disk magnets are affixed on the outer elements and on the inner resonator to introduce memory-wave interaction. (D) Numerical model of the module. The inter-layer magnetic coupling is modelled as linear spring foundations, a tensional spring $k_l$ and a rotational spring $k_\theta$, connecting each inner resonator with fixed ground. (E) Front and side view of the 11-by-11-unit experimental prototype.

With the proposed platform, the state-dependent magnetic coupling reconfigures the local resonances and, consequently, the phononic properties of the inner metamaterial plate, providing the crucial link between stored memory and wave physics. To this end, we investigate the programmable wave physics of the metastructure under different memory states via finite element simulations and experiments. Here, we consider small-amplitude out-of-plane elastic wave propagation, and a numerical model is established for simulation by modelling the magnetic coupling from the outer layers as equivalent linear spring foundations that connect each inner-layer resonator with fixed ground, including a tensional spring in the z-direction and a rotational spring about the x and y-axes (Figure 2D). The effective translational and torsional stiffnesses $k_l$ and $k_\theta$ under different memory states are characterized through experimental measurements and numerical magnetic modelling. Under state '1', the coupling yields a positive translational stiffness of $2.5 \times 10^3$ N/m and a negative torsional stiffness of $-0.011$ N·m/rad, whereas in state '0', both effective stiffnesses are negligible (see Supporting Information, Section S3, for details of the characterization of the magnetic coupling).

We first analyze the linear dispersion relation considering a basic periodic unit, where the Floquet-Bloch periodic boundary conditions are applied. The Bloch wavevector $\boldsymbol{k} = (k_x, k_y)$ is swept along the edges of the 2D irreducible Brillouin zone (shaded area in Figure 3D) within the reciprocal lattice. The unit's band structures under state '0' and '1' are obtained, as plotted in Figure 3A and 3B. In the considered frequency range from 0 to 600 Hz, the '0' state exhibits two complete local-resonant bandgaps for the out-of-plane modes: one between 256 Hz and 339 Hz, and another above 566 Hz. For the '1' bit state, the increased magnetic coupling significantly alters the band diagram. The lowest out-of-plane branch, representing translational motion in the z-direction (Figure 3C, star), shifts upward due to the positive translational stiffness, creating a new foundation-induced bandgap from 0 to 168 Hz. Concurrently, the other two out-of-plane branches, associated with rotational motion about the in-plane axes (Figure 3C, diamond/triangle), shift downward due to the negative torsional stiffness. Experimental measurements validate this memory-dependent band structure tailoring, showing strong agreement with these results (see Supporting Information, Section S4 for comparison of experimental measurement and numerical band structures).

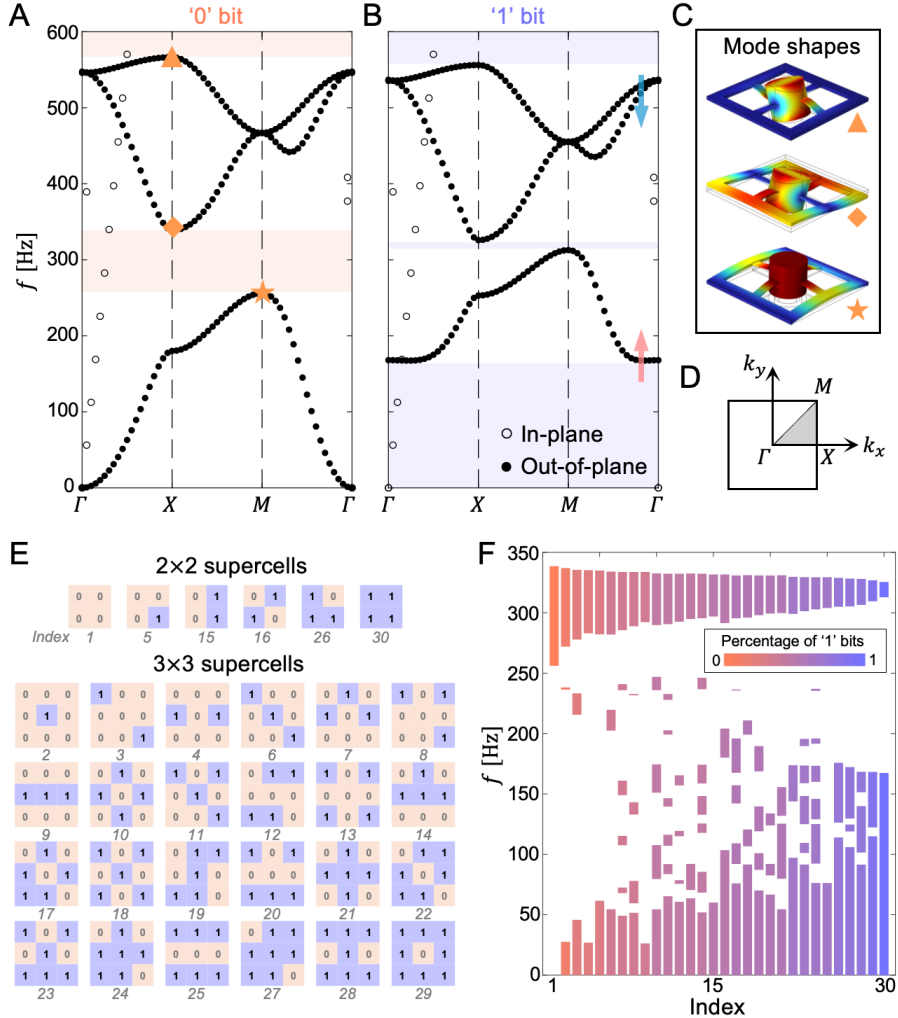

**Figure 3. Memory-programmed band structures.** (A, B) Band structures for a periodic unit under '0' (A) and '1' bit states (B). Circles indicate in-plane modes (x-y-plane vibration), and filled dots represent out-of-plane modes (z-direction). Shaded regions highlight identified out-of-plane bandgaps. (C) Mode shapes for the three out-of-plane branches. The lowest branch indicates translational modes (star), and the other two branches are rotational modes (triangle/diamond). (D) 2D irreducible Brillouin zone within reciprocal lattice (shaded area). The wavevector $(k_x, k_y)$ are swept along the edges of the irreducible Brillouin zone ($\Gamma \to X \to M \to \Gamma$). (E) 2-by-2 and 3-by-3 supercell memory configurations. 30 independent configurations are identified accounting for symmetry and periodicity equivalence. (F) Bandgaps across all configurations. The horizontal axis represents configuration indexes, and the vertical bars show the out-of-plane bandgap frequency regions for each configuration. The bar color encodes the fraction of '1' bits within each memory configurations.

We then generalize the analysis to periodic supercells by considering different spatial memory bit distributions within 2-by-2-cell and 3-by-3-cell units, which leads to more diverse wave behaviors. Accounting for periodicity and symmetry-induced equivalence, we identify 30 independent bit configurations (Figure 3E). For each configuration, the corresponding bandgap frequency regions are calculated, as summarized in Figure 3F, where the horizontal axis represents the indexes of different memory configurations and the vertical columns show the corresponding bandgap regions. We observe rich memory-dependent wave characteristics where each configuration produces a unique band diagram. Even with identical percentages of '0'/'1' bit ratio but different spatial distributions, the band diagrams can be entirely distinct, which can generate additional band branches across different frequencies. Physically, different arrangements of mechanical bits modify local resonances and couplings between units, which leads to lattice symmetry breaking and branch degeneracies splitting in diverse ways (see Supporting Information, Section S5 for more details of supercell dispersion analysis). This establishes a predictable and deterministic mapping between the encoded memory states and the resulting wave-propagation landscape. Fundamentally, it converts the memory into rich phononic information carried by the wave transmission spectrum that specifies the frequency channels conveying the stored information.

**Frequency-selective wave energy localizations for memory-pattern feature perception.** With the fundamental wave physics uncovered, we now investigate to discover how these memory-programmed phononic properties enable a more sophisticated capability: frequency-selective wave energy localization that performs feature-level perception of stored information. While previous research has exploited rich waveguiding behavior in phononic metamaterials [36,41–44], such as defect-induced energy confinement along specific paths or topological-symmetry-breaking-induced edge or corner modes, we uniquely utilize these phenomena to realize a mechanical perception system where the wave energy localization modes selectively perceive and extract informative geometric features of the stored memory patterns.

Here, we exploit defect-induced wave energy localization, where the 'defect' comes from the mechanical bit reconfiguration under information writing. We first uncover the fundamental principle through dispersion analysis on a strip supercell containing a single programmed defect state. When a single unit within a uniform background is switched to a different memory bit, it creates a local defect with resonance perturbation that traps wave energy locally at specific frequencies. For instance, in a strip of uniform '0' bits, programming the central unit to '1' bit (referred to as Type I encoding) introduces sharp defect modes within the original bandgap of the corresponding defect-free strip, as indicated by red dots in Figure 4A, left, concentrating vibrational energy precisely at the defect unit. Conversely, in a strip of uniform '1' bits, programming the central unit to '0' (referred to as Type II encoding) generates defect modes within the foundation-induced bandgap (Figure 4A, right), with similar spatial localization but at entirely different frequencies. The existence of these frequency-selective defect modes is further validated through full-scale metastructure experiments and simulations, as shown in Figure 4B, where different-frequency waves can be selectively guided and localized along defected paths by reconfiguring the metastructure memory states at excitation frequencies predicted by the eigen-frequency ranges of these defect-mode bands. This frequency-selective localization provides a direct mechanism for processing the stored information: defect modes act as spatial phononic filters, producing locally enhanced responses at the programmed sites so that patterns encoded in the memory can be read from the scattering wavefields.

This defect-mode physics scales elegantly to complex 2D patterns, enabling rich feature-level perception. As illustrated in Figure 4C, when encoding complex spatial information, such as letters '*U*' and '*M*' into the metastructure memory, the system generates a set of distinct and diverse defect modes that naturally decompose the patterns into constituent geometric features. Each defect mode localizes wave energy at specific locations, such as edges, sharp corners, or junctions, with the spatial wave energy distribution directly mapping to particular geometric characteristics of the encoded memory pattern. Crucially, these feature-modes resonate at distinct frequencies due to variations in local magnetic coupling and resonator interactions, which enables frequency-selective feature perception through spectral probing. Figure 4D presents the frequency response spectrum under the two memory patterns by applying harmonic excitation at the central unit. It is observed that the '*U*' pattern exhibits resonance peaks where its specific defect modes activate, while the '*M*' pattern displays a different set of peaks corresponding to its unique modes. Specifically, this spectral response separation enables pattern discrimination by probing the metastructure energy output at specific excitation frequencies. For example, as shown in Figure 4D, right, a 99 Hz excitation activates one of the 'U' pattern feature modes, resulting in clear energy-localized scattering wavefield with high energy output, while the 'M' pattern is under off-resonance with much lower energy output. Yet, the reverse occurs at 126 Hz (resonance for 'M'), indicating the feasibility of harnessing the wave responses to uniquely identify and classify different stored information. We note that, in these two cases, the information is encoded as Type II defects with '1' bits forming the patterns on a '0' background. Supporting information S6 provides more examples of feature perception under both Type I and Type II encoding.

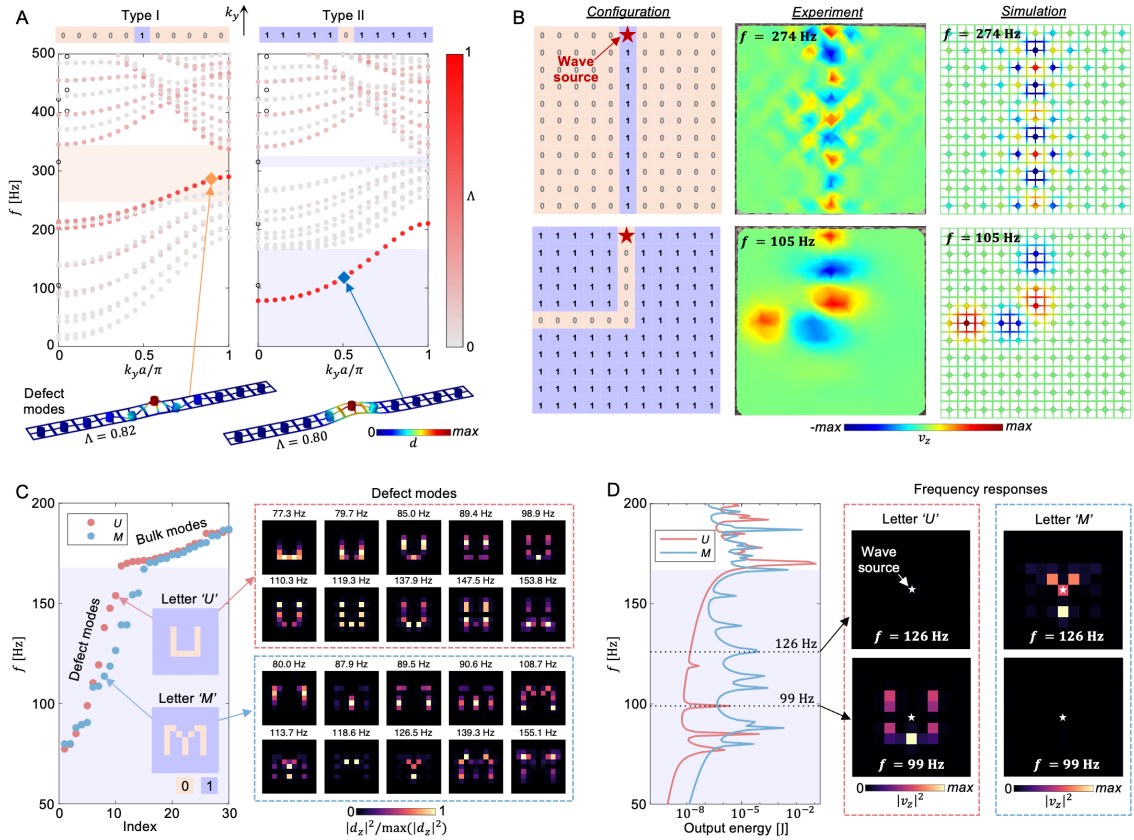

**Figure 4. Frequency-selective wave energy localization for feature perception.** (A) Band structures of a strip supercell with a central defect state under Type I and Type II encoding. The wavevector $k_y$ is swept from 0 to $\pi/a$, where $a$ is the unit cell length. Shaded regions highlight the bandgaps of the corresponding defect-free strip. A localization parameter $\Lambda = \iiint_{V_{defect}} d^2 \, dV / \iiint_{V_{total}} d^2 \, dV$ is defined to quantify the energy confinement at the defect unit, where $d$ is the displacement, and $V_{defect}$ and $V_{total}$ represent volume of the central unit and the strip, respectively, with $\Lambda \approx 1$ indicating defect modes (red bands). (B) Localization-mode-induced waveguiding in the metastructure through experiments and simulations. Two waveguides are programmed: a straight line of '1's in '0' background and an L-shaped '0' path in '1' background. At the defect-mode frequencies predicted in (A), steady-state out-of-plane velocity fields $v_z$ show energy localized and guided along the encoded paths. (C) Eigenfrequencies (left) and mode shapes (right) of the metastructure with letters 'U' and 'M' encoded as Type II defects. In the left plot, the x-axis represents mode index, and the y-axis indicates eigenfrequency. Mode shapes are shown as squared out-of-plane displacement field $|d_z|^2$, normalized for visualizing energy distribution within the metastructure. Each encoded pattern generates diverse and frequency-selective defect modes that decompose the shapes into small-piece geometric characteristics, enabling feature-level perception. (D) Simulated frequency responses under harmonic excitations at the central unit. In the left plot, the y-axis represents excitation frequency, and the x-axis is the metastructure's steady-state energy output. The right plots show scattering wavefields (squared out-of-plane velocity $|v_z|^2$) for the two patterns at 99 Hz and 126 Hz that selectively excite the corresponding defect modes and enable discrimination by probing wave responses. For each pattern, the scattering wavefields share the same color-bar scale.

As such, the phononic metastructure functions as a mechanical perception system in the frequency domain, where each vibrational localization mode acts as a specific feature detector, with the spatial distribution of wave energy indicating what geometric characteristic is being perceived and the corresponding resonance frequency providing the access channel for this feature. This transformation of spatial memory patterns into identifiable phononic features through physically grounded wave localization establishes the foundation for the proposed in-memory phononic learning framework.

**In-memory phononic learning for pattern recognition.** Building upon the feature perception capabilities, we create an in-memory phononic learning framework that enables the metastructure to perform general pattern classification tasks, as illustrated in Figure 1B. The core idea is to treat the fixed, memory-programmed metastructure as a physical feature extractor in the frequency domain, and learn only the input waveforms that selectively activate and probe the most class-relevant feature modes. The decision is then read directly from the metastructure output energy carried by these excitations.

For an N-class classification task, we define N input channels $\{I_1, \dots, I_N\}$. Each channel corresponds to one class and is a trainable linearly weighted combination of different harmonic tones, in form of $I_i(t) = \sum_{j=1}^{K} w_{ij} \sin(2\pi f_j t)$, where $\{f_1, \dots, f_K\}$ are sampled excitation frequencies that cover the frequency regions where the corresponding defect modes occur, and $\{w_{ij}\}$ are trainable parameters that determine the contribution of each frequency components. For a given encoded memory pattern, we measure the metastructure's steady-state response at each $f_j$ under the same excitation amplitude and compute an energy feature $E(f_j)$, i.e., the metastructure's output energy under excitation frequency $f_j$ in terms of summed vibration energy over all resonators. Considering small-amplitude linear regime, the time-averaged output energy under channel $I_i$ can be calculated as $E_{out}^{(i)} \propto \sum_j w_{ij}^2 E(f_j)$, which is the nonnegative linear combination of energy features under a constant scaling factor across channels. We then normalize these energy outputs across all channels to obtain an N-dimensional probability vector $p = [p_1, \dots, p_N]$, where $p_i = \frac{E_{out}^{(i)}}{\sum_{k=1}^{N} E_{out}^{(k)}}$. The class with the highest probability indicates the prediction. In practice, this probability vector is directly inferred by probing the energy output across the channels, where highest-energy channel indicates the predicted class. We train input waveform weightings $\{w_{ij}\}$ with cross-entropy loss on labeled training samples, where each of their energy features under harmonic excitations are pre-computed as training data. $L_1$ regularization is applied to promote sparse frequency selection, automatically identify the most discriminative feature modes for each class so that each learned input channel selectively activates the localized vibrational modes associated with its target class while suppressing response for other classes. Full details of the framework formulation, training process, and implementation are given in Supporting Information, Section S7.

In principle, while our framework is inspired by the architecture of PRC at a high-level by implementing a wave-physics-informed analog neural network, as depicted in Figure 1C, it is significantly and fundamentally different. Here, the fixed frequency responses of the metastructure serve as the physical reservoir, nonlinearly mapping an encoded memory pattern into a high-dimensional phononic feature space that inherently operates in parallel across frequencies. The trainable input waveforms then act as a linear output layer, combining these perceived features to perform classification. While standard PRC often relies on disordered systems to generate complex nonlinear dynamics needed for a rich feature space, our system uses rich and predicable wave physics from a structured metastructure medium. This *crucial difference* ensures that the high-dimensional feature mapping *emerges from physically traceable and interpretable localization modes*, rather than from the opaque, black-box dynamics. *More importantly, we realize learning and decision entirely within the phononic domain, which eliminates the need for the digital or analog electronic readout layers that are requisite in PRC, moving toward to truly self-contained cognitive mechanical intelligence.*

*Task 1: Geometric pattern classification.* To validate the in-memory phononic learning framework, we first pursue a proof-of-concept task to classify geometric pattern shapes—squares, triangles, and crosses—encoded into the metastructure memory. These patterns are randomly generated with diverse scales and locations (Figure 5A and 5E). We evaluate both Type I (pattern as '1's) and Type II (pattern as '0's) encoding schemes using a dataset of 150 samples (50 per class of pattern), with 20% samples reserved for testing. The framework achieves 100% classification training and testing accuracy for both encoding types in simulation, as indicated in Figure 5C and 5G, where the output energy profiles from the well-trained input channels under each sample are plotted, revealing that the highest energy outputs always come from the targeted channel. To gain more physical insight, Figure 5B and 5F plot the frequency component weightings from the learned input channels for each case, as well as the corresponding mean frequency response among each class of patterns. It is noticed that, for each input channel, those most significant frequency components with higher weightings concentrate on well-isolated resonance peaks for the targeted class while being off resonance for the other two classes. It reveals a *physics-grounded strategy* of the proposed framework, where the training selects a sparse set of frequencies at which each input wave only activates the corresponding class-specific localization modes while suppressing the modes associated with the completing classes, and as a result, the output energy across channels is redistributed with optimally enhanced vibration energy level at the corrected channel. This observation is further confirmed by examining the output wavefields of the metastructure under the well-trained input waves. For instance, Figure 5D and 5H visualize the scattering wavefields under the three well-trained channels for selected samples. We observe that $I_1$ activates square-related modes with energy localization at the diagonal corners of squares, while $I_2$ activates triangle-related modes along the triangle's inclined edges, and $I_3$ selects cross-related modes concentrating energy at the intersection region. This demonstrates that the learning process automatically identifies and leverages the most geometrically discriminative features for classification. Experimental validation of the geometric pattern classification task is shown in Figure 6, using a set of 54 training and 18 testing samples under Type I encoding. Figure 6A presents the measured steady-state wavefields for representative samples at six excitation frequencies that selectively activate each class-related feature modes, revealing the perception capability of the testbed. Figure 6B plots the normalized output energies across the trained input channels for all samples, where 53 out of 54 training samples and all testing samples are correctly recognized despite experimental noise, validating the effectiveness of the proposed framework. (see Supporting Information, Section S8 for more details of numerical and experimental validations of geometric pattern classification tasks).

Furthermore, by comparing the results from Type I and Type II encoding schemes, we find that the quality of wave localization directly impacts learning efficiency and confidence. Although both achieve 100% accuracy, we quantify the confidence of the results by the average normalized energy in the targeted channels across all samples, which is equal to the sample-averaged predicted probability of the true class. Type II encoding, which produces strong, well-separated defect modes within a wide foundation-induced bandgap, requires only 8 frequency components across all input channels and leads to a high confidence of 0.985, indicating that nearly all vibration energy concentrates in the correct channels with minimal leakage to the untargeted ones. By contrast, for Type I encoding, the defect modes appear in narrower local resonant bandgap associated with the state '0', where band overlapping and modal coupling with boundary modes are observed (see Supporting Information, Section S8 for more detailed analysis). This results in less distinct and noisier feature modes. Consequently, achieving the same accuracy requires a greater number of frequency components, and the output energy shows more diffusion into untargeted channels, yielding a lower mean normalized energy of 0.880. This underscores that a clearer physical mapping between memory and wave localization leads to more robust and efficient learning.

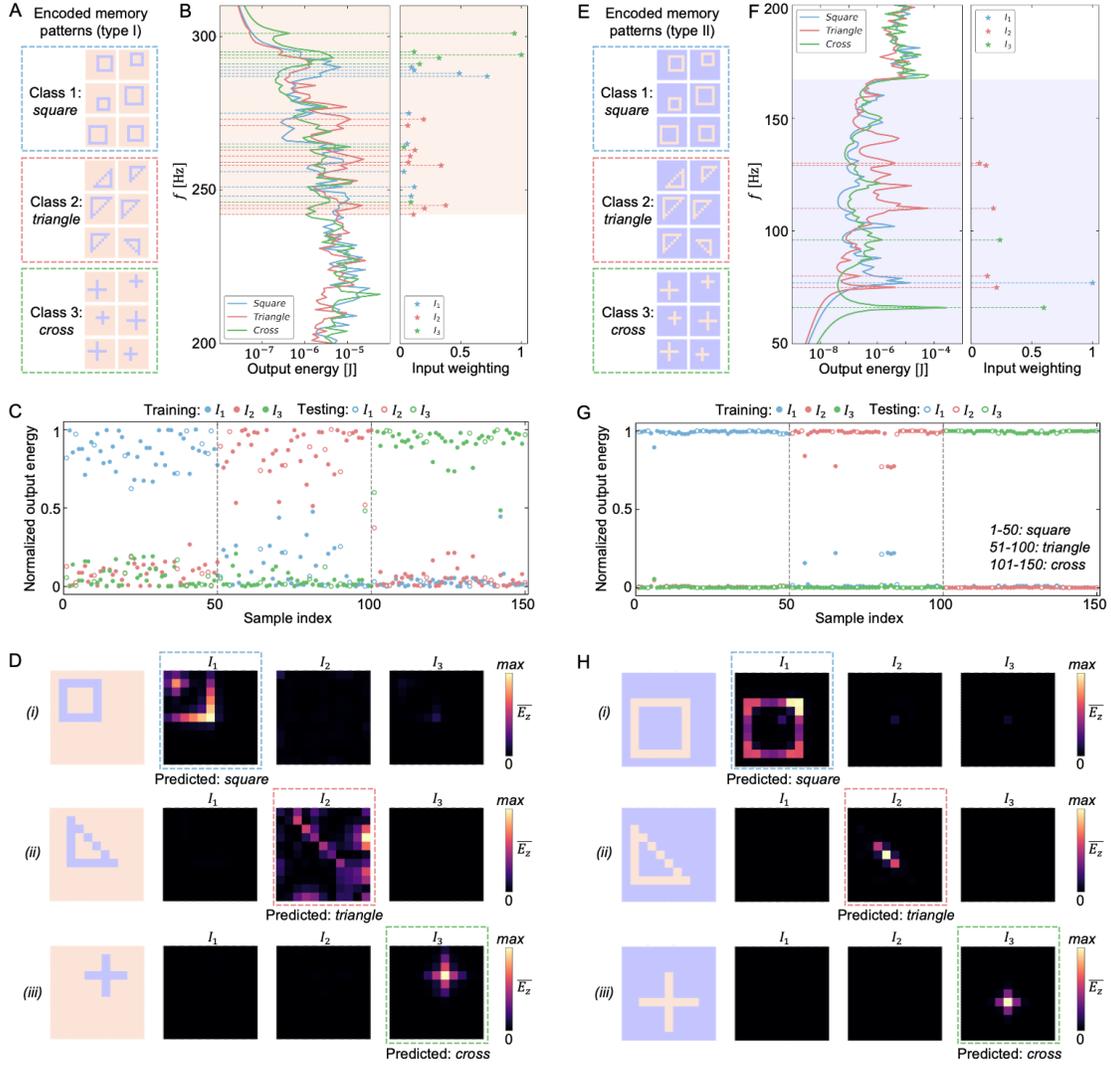

**Figure 5. Geometric pattern classification task.** (A,E) Example encoded patterns for three classes (square, triangle, cross) under Type I (A) and Type II (E) encoding. (B,F) Mean frequency response (left) for each class of patterns and the learned input waveform weightings (right). Dominant weights align with well-isolated class-specific resonance peaks and remain off-resonance for untargeted classes, revealing a sparse, interpretable frequency-selection strategy. (C,G) Normalized output energy for all samples under the three input channels, with training and testing partitions indicated. Samples 1-50 correspond to squares, samples 51-100 are triangles, and samples 101-150 are crosses. Both encodings reach 100% simulation accuracy for both training and testing scenarios; Type I achieves lower classification confidence (more energy diffusion into untargeted channels) with 25 frequency components selected, while Type II achieves higher confidence with only 8 frequency components due to cleaner defect modes in the wide foundation-induced gap. (D,H) Scattering wavefields (steady-state time-averaged energy response) for selected samples under the trained channels show energy localization at class-specific feature modes: diagonal-corner modes for squares under $I_1$, inclined-edge modes for triangles under $I_2$, and intersection modes for crosses under $I_3$. For each sample, the scattering wavefields share the same color-bar scale.

*Task 2: Handwritten digit recognition task.* To investigate generalizability, we apply the same metastructure to a more complicated and challenging task: classifying handwritten digits '0' and '1' from the benchmark MNIST dataset[45]. We select 150 samples per digit, down-sampled from 28-by-28 greyscale images to 11-by-11 binary patterns for encoding into the metastructure (Figure 7A), with 20% samples reserved for testing. Using Type II encoding, the system achieves a training accuracy of 98.3% and a testing accuracy of 93.3% in simulation by simply retraining the input waveforms (Figure 7B and 7C). Representative scattering wavefields under the trained input channels (Figure 7D) reveal the underlying physics: $I_1$ activates loop-closing modes characteristic of digit '0', localizing energy along a ring-like path (or a substantial arc), $I_2$ whereas excites edge-following modes aligned with the shape of a digit '1'. Despite substantial intra-class variation in thickness, curvature, and centering for handwritten digits, the classifier maintains strong performance in perceiving theses class-specific feature modes, where 37 frequency components across channels are used in this task to facilitate shifts of defect-mode frequencies among different samples. The results demonstrate the multi-tasking capability of the framework, where the hardware remains unchanged and only the input excitations are updated—a significant practical advantage over conventional morphological approaches that often require laborious, task-specific structural optimization and reconfiguration across different tasks. Furthermore, it demonstrates a more efficient pathway to scale computational complexity. Conventional morphological systems typically sequentially cascade multiple physical layers to enrich computing power, which increases architectural complexity and can compromise efficiency. In contrast, our framework scales within a single metastructure substrate by leveraging a richer set of frequency components. This essentially creates a high number of parallel, hidden layers in the frequency domain, where different frequency bands simultaneously extract diverse features. This inherent parallelism allows our system to handle more complex tasks like digit recognition without sacrificing its streamlined, single-layer architecture, while achieving high accuracy.

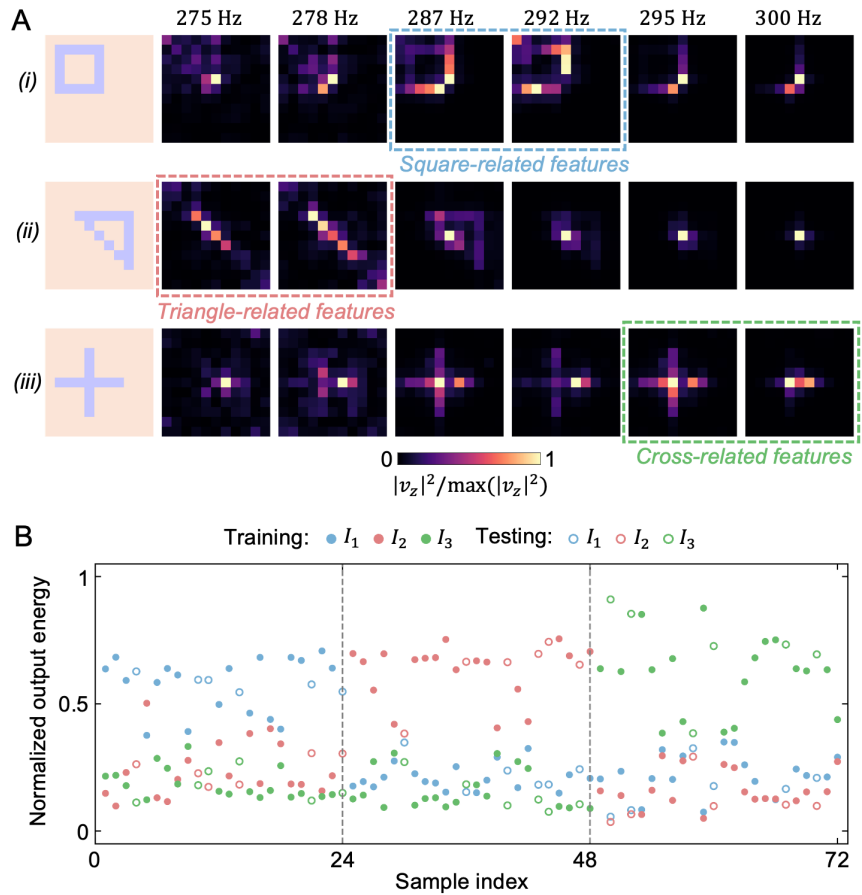

**Figure 6. Experimental validation of geometric pattern classification task.** (A) Measured steady-state responses (normalized squared out-of-plane velocity field $|v_z|^2$) of representative samples under six excitation frequencies, showing frequency-selective wave energy localization that activates class-related feature modes. (B) Normalized output energy for all samples (Type I encoding) under the three input channels, with training and testing partitions indicated. Samples 1-24 correspond to squares, samples 25-48 are triangles, and samples 49-72 are crosses. 53 out of 54 training samples and all 18 testing samples are correctly classified.

To further assess the information processing capacity of our platform, we extend the '0'/'1' handwritten digit classification tasks to a 14-by-14 metastructure in simulation. With increased spatial resolution and a larger dataset with 900 training and 100 testing samples per digit, the performance improves with reduced overfitting, achieving a training accuracy of 95.1% and testing accuracy of 96.0% (see Supporting Information, Section S9 for more details of handwritten digit recognition on a 14-by-14 metastructure). We then challenge the system with a full 10-class digit recognition task. For this highly complex task, we focus on highlighting the computing potential of the wave-based feature perception capability by employing a hybrid setup, where the frequency response serves as a fixed physical reservoir to perceive input pattern features, and for each harmonic excitation, the full spatial phononic wavefield is readout and concatenated as a feature vector. A linear readout layer is then trained to classify the resulting high-dimensional feature vectors. With 9000 training and 1000 testing samples, this setup achieves 96% testing accuracy, demonstrating that the phononic feature space perceived by the metastructure is sufficiently rich and discriminative for state-of-the-art tasks (see Supporting Information, Section S9 for more details of the 10-class handwritten digit recognition task).

The results from the two exemplar tasks collectively demonstrate that our in-memory phononic learning framework successfully leverages the rich wave physics of structured media to achieve machine-learning classification tasks directly in the mechanical domain in an efficient and physically interpretable manner. More specifically, by programming wave-memory interactions in a single metastructure, we enable a form of mechanical cognition, where the material itself can store and perceive information, learn from knowledge, and make decisions autonomously. This approach clearly establishes a new paradigm for creating intelligent matters capable of adaptive reasoning and learning through their inherent physical properties.

**Discussion**

In summary, we present an efficient, unified and physically transparent framework for advancing the state of the art of mechanical intelligence – embodying *in-memory phononic learning* into architected materials that enables memory, perception, and learning capabilities, and investigate its efficacy in and through a rationally designed phononic metastructure. By coupling non-volatile mechanical memory with wave-based information processing, we enable mechanical learning by harnessing memory-state-dependent wave dynamics as a physical analog neural network. It achieves effective feature-level perception through interpretable wave-localization modes and implements cognitive memory-pattern classification via trainable input waveforms, completing the entire process—from information encoding to perceptual feature extraction to learned decision— all within the mechanical domain. Utilizing example testbeds, our experimental and numerical validation demonstrates the framework's effectiveness, where the same metastructure achieves high accuracy in classifying geometric shapes and handwritten digits, with multi-tasking capability enabled simply by updating the input waveforms. This approach overcomes key limitations in the state-of-the-art mechanical intelligence. First, it provides physical transparency—unlike black-box morphological systems, the learned mappings are directly traceable to physics-grounded phononic property tailoring. Second, it offers high efficiency through phononic parallel processing across frequencies within a single material layer, avoiding the need for complex cascading of basic logic modules or layers of physical substrates. Third, it possesses versatile computing power and generalizability to diverse tasks, eliminating the need for task-specific structural optimization or reconfiguration.

Ultimately, this research shifts the ambition from creating *materials that merely compute* to engineering *cognitive matter* that can sense, reason, learn and understand through its intrinsic physical dynamics. These advances would support real-world applications including autonomous payload identification in adaptive infrastructures, tactile sorting in soft robotics, and embedded structural health monitoring where the material itself diagnoses damage patterns, opening new frontiers for the next generation of autonomous engineering systems with lower power consumption, more direct interaction with surroundings, much better cybersecurity, and greater survivability in extreme environments. The core principles demonstrated here—using structured wave physics for in-memory learning—are transferable across physical systems, such as photonic crystals for optical computing[46,47], and acoustic devices for sonic communication and processing[48]. Furthermore, this work offers an efficient, inherently parallel and transparent hardware platform for physical embodiment of machine-learning-classification tasks, which could inspire new wave-based machine learning architectures with enhanced interpretability.

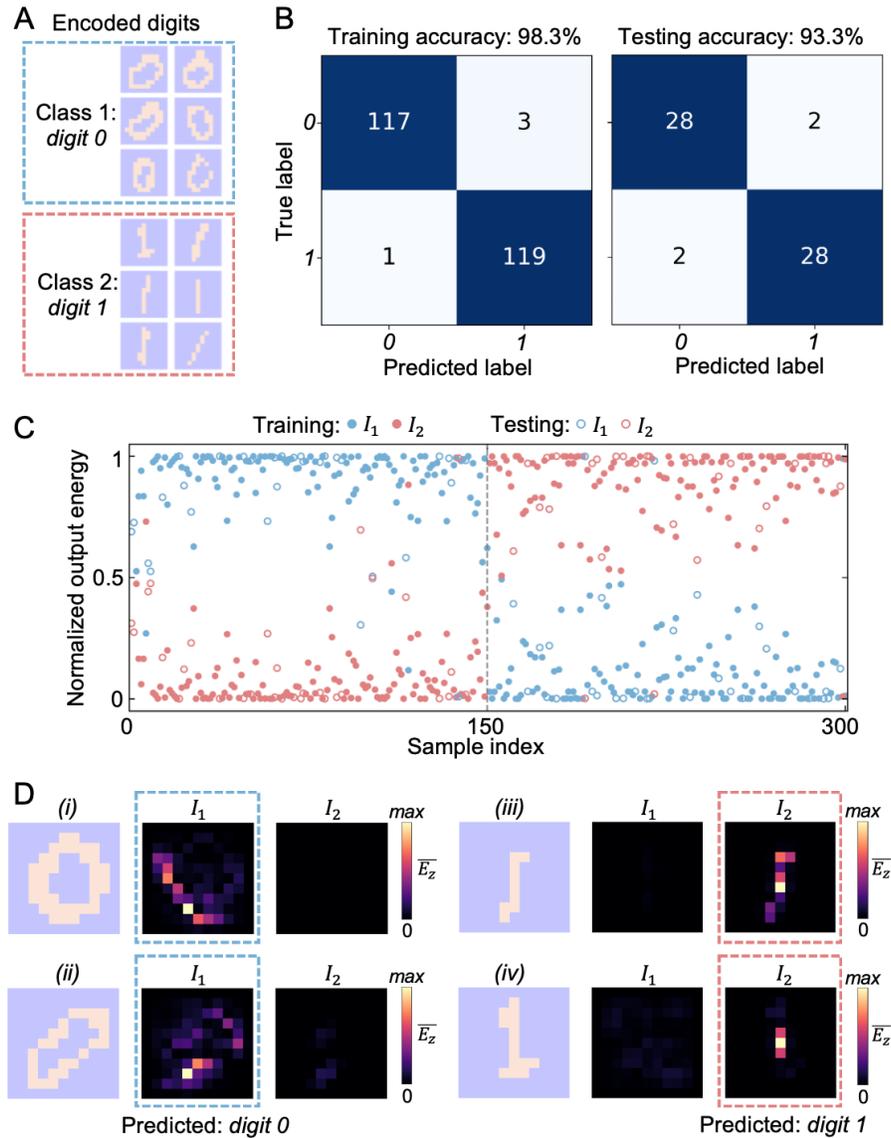

**Figure 7. Handwritten digit recognition tasks using the same metastructure.** (A) Example down-sampled digits '0' and '1' encoded as Type II patterns. (B) Training and testing confusion matrices. For each class, 120 samples are selected for training, and 30 samples are used for testing. 98.3% training accuracy and 93.3% testing accuracy are achieved. (C) Normalized output energy across samples for the two trained channels. Samples 1-150 correspond to digit 0, and samples 151-300 are digit 1. The predicted label corresponds to the channel with maximal energy. (D) Scattering wavefields under trained input waveforms for representative test digits. The '0' channel ($I_1$) excites loop-closing localization around a ring-like path, while the '1' channel ($I_2$) excites edge-following modes, yielding separability between digits '0' and '1'. For each sample, the scattering wavefields share the same color-bar scale.

**Materials and Methods**

*Metastructure prototype fabrication and experimental testing.* An 11-by-11 multistable phononic metastructure is built by multi-material 3D printing and modular assembly. Full geometry, materials, and fabrication details are provided in Supporting Information Section S1. Quasi-static tensile testing is performed on an Instron machine to validate module multistability and quantify inter-layer magnetic coupling. Wave transmission characteristics and experimental training data are collected using a scanning laser Doppler vibrometer with an electromagnetic coil shaker providing excitation. A detailed description of experimental setup is provided in Supporting Information Section S2.

*Simulation.* Finite element analysis is conducted in COMSOL Multiphysics 6.2. The magnetic coupling modelling is conducted using AC/DC module, and all other simulations are conducted with Structural Mechanics model. The eigen-frequency studies are performed to obtain unit and supercell dispersion, and the frequency-domain studies are conducted to obtain frequency responses and collect simulation dataset of the classification tasks. Details of simulation setups can be found in Supporting Information Section S3-S6.

*Learning framework.* The phononic learning framework is implemented with PyTorch 2.0.1 in Python 3 environments. The learning ratio and regularization parameter are adaptively tuned and searched via grid search to enhance testing accuracy for each task. Detailed setup can be found in Supporting Information Section S7-S9.

**Supporting Information**

Supporting Information, including detailed methods and supplementary figures, will be made available upon formal publication

**Acknowledgments**

The authors acknowledge the financial support of the Air Force Office of Scientific Research under Award No. FA9550-23-1-0466, and partial support from the National Science Foundation under Award CMMI 2328523.